# Broadband Four-Wave Mixing Enhanced by Plasmonic Surface Lattice Resonance and Localized Surface Plasmon Resonance in an Azimuthally Chirped Grating

Abhik Chakraborty,[†,‡] Parijat Barman,[†,‡] Ankit Kumar Singh,[‡] Xiaofei Wu,[‡] Denis A. Akimov,[‡] Tobias Meyer-Zedler,[†,‡] Stefan Nolte,[∥,§] Carsten Ronning,[⊥] Michael Schmitt,[†] Jürgen Popp,[†,‡,*] Jer-Shing Huang[†,‡,#,††,*]

[†]Institute of Physical Chemistry and Abbe Center of Photonics, Friedrich-Schiller-Universität Jena, Helmholtzweg 4, D-07743 Jena, Germany

[‡]Leibniz Institute of Photonic Technology, Albert-Einstein Str. 9, 07745 Jena, Germany

[∥]Institute of Applied Physics, Abbe Center of Photonics, Friedrich-Schiller-Universität Jena, Albert-Einstein-Straße 15, Jena 07745, Germany

[§]Fraunhofer Institute for Applied Optics and Precision Engineering IOF, Center of Excellence in Photonics, Albert-Einstein-Str. 7, Jena 07745, Germany

[⊥]Institut für Festkörperphysik, Friedrich-Schiller-Universität Jena, Max-Wien-Platz 1, 07743 Jena, Germany

[#]Research Center for Applied Sciences, Academia Sinica, 128 Sec. 2, Academia Road, Nankang District, Taipei 11529, Taiwan

[††]Department of Electrophysics, National Yang Ming Chiao Tung University, Hsinchu 30010, Taiwan

**ABSTRACT:** Plasmonic enhancement of nonlinear light-matter interaction can be achieved via dedicated optimization of resonant plasmonic modes that are spectrally matched to the different wavelengths involved in the particular nonlinear optical process. In this work, we investigate the generation and enhancement of broadband four-wave mixing (FWM) in a plasmonic azimuthally chirped grating (ACG). The azimuthally varying grating periodicity in an ACG offers a well-defined channel to mediate the near field and the far field over a broad range of wavelengths. However, the particular mechanism responsible for field enhancement in such a platform depends on the interplay between the effects manifested by both the groove geometry and the grating's periodicity. This work delineates the collective contribution of groove geometry-dependent localized surface plasmon resonance (LSPR) and periodicity-dependent plasmonic surface lattice resonance (PSLR) over a broad range of wavelengths to bring into effect the enhancement of broadband FWM in an ACG.

## 1. Introduction

Nonlinear signal generation stems from the nonlinear variance of induced polarization with the incident electromagnetic field via the higher-order corrections to the susceptibility in the matter.[1] Four-wave mixing (FWM) is a third-order nonlinear optical process. Its intriguing properties can be harnessed to manifest applications in biosensing,[2] nonlinear microscopy, and imaging,[3] telecommunications,[4] all-optical switching,[5] signal regeneration,[6] phase-sensitive amplification,[7] entangled photon pair generation[8], and metrology.[9] Enhancement of nonlinear light-matter interaction by carefully tailoring the geometry of nanometer-scale cavities is a very effective way to amplify these processes.[10-13]

Rationally engineered plasmonic geometries can provide spatio-spectral control over localized plasmonic modes. When the localized near field is resonant at every input and output frequency relevant to the nonlinear optical process, these structures become efficient in facilitating nanostructure-enhanced nonlinear light-matter interaction.[14-17] It is well known that a plasmonic hot spot is endowed with a steep field gradient in a tightly confined space. This is accompanied by a broadband momentum distribution in reciprocal space.[16] Therefore, a plasmonic hot spot serves as an efficient agent for electromagnetic field enhancement and nonlinear frequency conversion. For a third-order nonlinear optical process like FWM that demands high input intensities and phase matching, the aforementioned attributes of plasmonic hot spots are important for the enhancement in efficiency pertaining to the required phase-matching and the FWM signal generation.

Plasmonic gratings are effective systems for coupling light from the far field into localized hot spots in the near field, and *vice versa*. In fact, linear plasmonic gratings have been used on multiple occasions to enhance FWM.[18-20] Similarly, a plasmonic azimuthally chirped grating (ACG) has been used to investigate the enhancement effect of plasmonic gratings in 2-color surface-enhanced coherent anti-Stokes Raman scattering (SECARS).[17] It was experimentally demonstrated that having a plasmonic grating resonant at the pump (input), Stokes (input) and the anti-Stokes (output) frequencies is crucial for enhancing the signal of 2-color CARS because the grating plays the role of a receiving and transmitting optical antenna for the input beams and the output signals, respectively.

The working principle of the ACG-enhanced nonlinear signal generation can be described with a three-step process. Firstly, the grating of the ACG helps to couple in the far field excitation at every relevant input frequency to the localized hot spots in the grooves. In this case, the ACG functions like a receiving optical antenna. Secondly, the steep field gradient and the corresponding broadband momentum distribution of the hot spots promote the nonlinear optical process. Finally, the generated output at the near field is coupled out via the grating's antenna effect again. The three-step description of the ACG's working principle has been an effective model, provided the optical response of the plasmonic grating as a function of the periodicity is well known. Figure 1a depicts the 3-step mechanism of the enhancement provided by the ACG.

In our previous works, the ACG works in the regime, where its optical response can be well described analytically by considering only the periodicity.[21-24] We showed that the ACG could be designed to have grating sections simultaneously supporting photon-to-surface plasmon polaritons (SPP) coupling at various frequencies relevant to the nonlinear optical process. The suitable sector areas of the ACG could be predicted solely by considering the photon-SPP momentum-matching condition. This provides a simple and effective model for the prediction of the best grating period for nonlinear signal enhancement. In addition to the periodicity-dependent regime, plasmonic gratings can work in the regime where the resonance of individual grooves dominates the optical response. For example, Wang et al. demonstrated that second harmonic generation from a silver plasmonic grating is mainly determined by the groove depth but rather insensitive to the periodicity.[25] In this case, the groove geometry is critical for optimizing the enhancement of nonlinear signal generation. In our ACG, this happens in the sections where the period is low and is not comparable to the wavelengths relevant to the nonlinear optical process. Overall, the optical response of a plasmonic grating, depending on the operational regime, can be dominated by the period-dependent grating response, the groove geometry-dependent local resonance, or both of them. To optimize the grating, multiple parameters, including the grating period and the groove geometry, should be scanned.[19] Compared to typical linear gratings, ACGs are a more convenient platform because they offer spatially and spectrally resolved broadband periodicity and thus include all these scenarios in one single platform. Therefore, an ACG offers multiple degrees of freedom for the optimization of the optical response for plasmonic enhancement at various input and output frequencies in one single platform. The ACG used in this work operates in a regime, where the optical response is mainly determined by the local resonance of individual grooves and its interplay with the period-dependent grating resonance.

As our ACG is an azimuthally chirped grating consisting of periodic V-shaped grooves in a gold film, the local resonance is the localized surface plasmon resonance (LSPR) in the vertical V grooves. The interplay of the LSPR of the grooves with the periodic lattice of the grating results in the so-called plasmonic surface lattice resonance (PSLR), which stems from the constructive interference of the LSPR of individual grooves coupled via the periodicity-dependent in-plane SPPs.[25-30] A general case of surface lattice resonance happens when the scattered light in the plane of an array of solitary plasmonic nanoresonators resonantly excites in-phase oscillation of the adjacent plasmonic nanoresonators.[26] SLR features a sharp Fano line shape along with a high quality factor due to the reduction of radiative loss.[26, 31] Typical plasmonic gratings are arrays of periodic V-shaped grooves on a metal film. Each groove can be considered as a vertical resonator for LSPR. When the in-plane SPPs launched by individual grooves reach the adjacent grooves in phase, collective oscillations of the LSPR in all grooves can be excited and the PSLR condition is fulfilled. ACGs contain broadband chirped grating periodicity. Therefore, depending on the operational frequency range, the optical response of the ACG can be dominantly determined by the groove's LSPR and the periodicity-dependent PSLR. In this work, we investigate the enhancement of broadband nonlinear signal generation by a broadband gold ACG. This is operated in the regime where the optical response of the ACG majorly reveals the LSPR of the groove and the PSLR from the coupling of the LSPR with the periodicity-dependent grating effect.

The broadband FWM demonstrated in this work is achieved with a broadband (BB) fs-pulsed laser and a narrowband (NB) ps-pulsed laser at the frequency $\omega_{NB}$. This allows for the generation of broadband signals simultaneously from 2-color and 3-color FWM schemes (Figure 1b). In the 2-color FWM scheme, the signal stems from the mixing of an NB photon with the instantaneous polarization in gold created by the coherent pumping of the structure by an NB photon and a BB photon. In a 3-color scheme, various combinations of three photons at different frequencies can lead to the same FWM signal. In this scheme, each combination consists of two photons at different frequencies (marked as $\omega_{BB}$ and $\omega'_{BB}$) from the BB pulse and one photon from the NB pulse. One example is given in the right panel of Figure 1b, where two photons at different frequencies from the BB pulse coherently produce the instantaneous polarization in gold at the frequency of $\omega_{BB} - \omega'_{BB}$, which then mixes with a photon from the NB to produce the FWM signal. In this case, the excitation of the instantaneous polarization is broadband in nature.

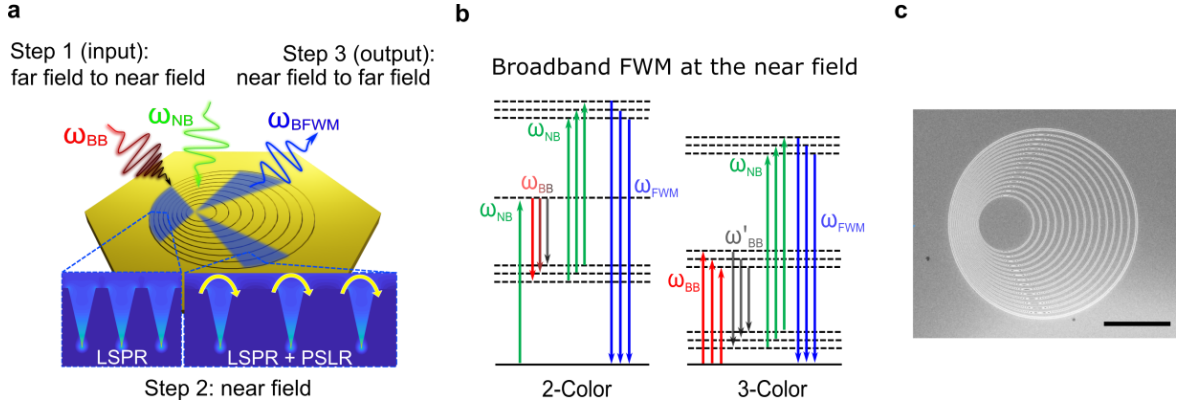

**Figure 1.** (a) Schematic illustration to describe the mechanism of surface-enhanced broadband FWM in gold ACG. The ACG works as an efficient in-coupling and out-coupling platform, as demonstrated in step 1 and step 3 respectively. Step 2 exhibits the intermediate near-field effect where the LSPR facilitates the phase-matching for broadband FWM and influences along with the PSLR the overall enhancement of the nonlinear light-matter interaction. The contribution from the PSLR at the spectral regime we are interested in has been exhibited in the periods that are comparable to the wavelength of the PSLR in the step 2 of the schematic. This replicates the real-world condition where the PSLR at the wavelength relevant to our work is pronounced only at certain periods of the ACG while the LSPR is pronounced as long as the groove geometry supports it. The lower periods are expected to exhibit higher response from the LSPR as the packing density of grooves within the FWM focal spot is larger there. (b) Energy diagram of the broadband FWM by the 2-color and the 3-color scheme. (c) SEM image of an ACG with fifteen rings on which the broadband FWM experiment has been conducted. Scale bar: 10 $\mu m$.

Since the phase-matching conditions between the different input frequencies and the output frequency for each combination of input and output in the 3-color process are satisfied by the broadband momentum of the plasmonic near field of the ACG's hot spots, we compare the experimental FWM signal distribution with the numerically simulated broadband optical near field in the grooves to understand the resonance responsible for the FWM enhancement. The hyperspectral investigation of the entire process allows us to demonstrate the broadband flexibility and the role of a plasmonic ACG in the enhancement of broadband FWM in a profoundly rigorous manner.

A schematic of the entire mechanism through which the aforementioned description of surface-enhanced broadband FWM takes place in the plasmonic ACG has been provided in Figure 1a. It must be emphasized that the dominant contributing modes in the enhancement mechanism have been characterized to be LSPR and PSLR in this work. Since the LSPR is governed predominantly by the groove geometry, it is ubiquitous in the ACG. The PSLR, on the other hand, becomes pronounced in the spatial regime of periods where the periodicity is comparable to the input and output wavelengths contributing to the FWM process.

## 2. Design, analysis and optical setup

### 2.1 Design and fabrication of the plasmonic ACG

The plasmonic ACG consists of a set of eccentric circular grooves in a gold film. The trajectory of the $n^{th}$ circular groove can be described as,

$$(x - nd)^2 + y^2 = (n\,\Delta r)^2 \qquad (1)$$

Here, $d$ is the displacement between the centers of two consecutive circular rings, whereas, $\Delta r$ is the increment in the radii of the rings. The grating periodicity ($P$) of an ACG is a function of the in-plane azimuthal angle ($\varphi$), which can thus be described as,

$$P_{\Delta r, d}(\varphi) = \left| d\cos\varphi \pm \sqrt{(d^2 \cos 2\varphi + 2\Delta r^2 - d^2)/2} \right| \qquad (2)$$

The ACG used for this work has been designed to have fifteen rings, where the smallest period is 300 nm ($\varphi = 180°$) and the largest period is 1100 nm ($\varphi = 0°$). To fabricate the designed ACG, focused Gallium-ion beam (NanoLab 600, FEI Helios) has been used to mill the circular grooves in an atomically flat monocrystalline gold flake with a thickness of ~255 nm obtained from chemical synthesis.[32] The fabricated ACG used in this work is shown in Figure 1c. The milled grooves have been estimated to have a depth of around 150 nm and an opening width of around 70 nm. Simulations are performed according to the experimentally measured geometrical parameters.

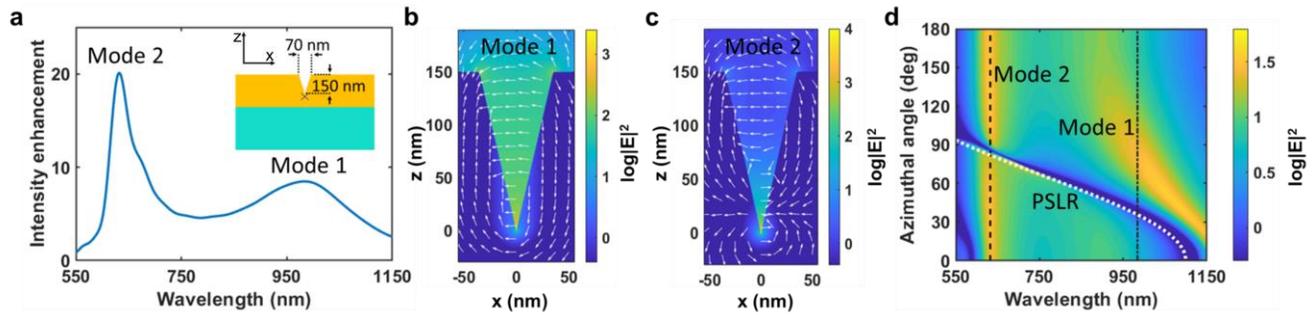

**Figure 2.** (a) Simulated near field response at a point monitor 3 nm below a single groove. The simulated result exhibits two separate peaks in the intensity enhancement titled Mode 1 and Mode 2 at wavelengths around 1000 nm and 625 nm, respectively. Inset: Schematic illustration of a single V-shaped groove with 150 nm depth and 70 nm opening width in a 255 nm thick gold layer with a point monitor marked as × positioned 3 nm below the groove. (b,c) Intensity distribution (logarithmic scale) and corresponding electric field vector profile are shown for Mode 1 and Mode 2 in an individual groove. (d) Near field response simulated at the point monitor for the infinitely extending series of V-shaped grooves with different periodicities. The different periodicities have been simulated to account for the near field response at different azimuthal angles (y-axis) of the ACG. A broadband spectral regime has been plotted on the x-axis. The peaks of the LSPR modes Mode 1 and Mode 2 for the single groove configuration from Figure 2a are marked with a black dot-dashed line and a black dashed line, respectively. The PSLR in the periodic system, which has been analytically calculated, has been marked with a white dotted line.

## 2.2. Simulating the near field response of the ACG

While the electronic resonances in gold are responsible for the locally generated FWM signal, the geometrical parameters of the nanostructure determine the resonances of the structure and thus the local optical field for the electronic oscillation responsible for the nonlinear optical process. Therefore, understanding the enhancement mechanism with respect to the geometry of our plasmonic ACG is pivotal for this work.

By using FDTD (Lumerical 2020 R2.3), we have simulated the responses from both an individual groove in gold and from periodic grooves in gold that represent the ACG. The source used in the simulation is a plane wave at normal incidence. A single V-shaped groove in gold with a groove depth of 150 nm and an opening width of 70 nm has been considered for our simulation of the single groove response. The near-field response has been calculated at a point monitor 3 nm below the groove. The simulated near-field spectrum for a single groove in gold with the aforementioned dimensions are shown in Figure 2a, where two distinct LSPR peaks (Mode 1 and Mode 2) can be observed. Mode 1 and Mode 2 are around 1000 nm and 625 nm, respectively.

From the corresponding near-field intensity enhancement and electric field vector profiles of the two modes (Figures 2b and 2c), it is clear that Mode 1 is the fundamental LSPR mode and Mode 2 is the higher-order LSPR mode. The LSPR modes are important for the enhancement of the input beams and the output signals in the FWM process. In addition to simulating the single groove response, we have also investigated the near field response of an infinitely extended array of such grooves with respect to the periodicity, i.e., the response of the plasmonic grating. Taking into account the azimuthal angle-dependent periodicity of our ACG, we have simulated the near-field response for different periods corresponding to different azimuthal angles (Figure 2d). The azimuthal angle-dependent periodicities in ACG is calculated by using Equation 2. The resonances of the single groove LSPR, i.e., Mode 1 and Mode 2 are marked by a black dot-dashed line and a black dashed line, respectively, in Figure 2d. As the in-plane azimuthal angle changes, i.e., the periodicity of the grating varies, a periodicity-dependent PSLR mode with a Fano-like spectral profile is observed in addition to the two LSPR modes. Although the groove geometry does not change with the grating period, the LSPR modes (Mode1 and Mode2) still shift with respect to the azimuthal angle because they couple to the periodicity-dependent PSLR. The appearance of the PSLR contributes to an avoided crossing region (white dotted line) throughout the investigated broadband range of wavelengths. The white dotted line has been analytically calculated to model the PSLR in the periodic system. It must be emphasized that delineating the nature of the influence of the dispersive permittivity of gold on the PSLR remains beyond the scope of this work. However, in our investigated spectral regime, the effective permittivity of the gold-air interface which is experienced by the SPP is very close to the permittivity of the dielectric surrounding. Therefore, the wavelength at which the PSLR comes into effect for a particular period $P$ can be analytically predicted (marked with the aforementioned white dotted line in Figure 2d) by,[33, 34]

$$\lambda_{\text{PSLR}} = \frac{Pn_d(1\pm\sin(\theta))}{m}, m = \pm1, \pm2, \ldots \quad (3)$$

Here, $n_d$ is the refractive index of the dielectric surrounding and $\theta$ is the angle of incidence. In our calculation, the dielectric surrounding has been considered to be air and the incidence for both our calculation and our simulation has been considered to be along the normal to the surface of the ACG. Along the white dotted curve, considerable quenching accompanied by the adjacent steep field enhancement can be seen in the specific combinations of azimuthal angles (periods) and wavelengths where the PSLR resonance with Fano lineshape comes into effect. Therefore, in certain combinations of period and wavelength, the FWM process undergoes PSLR-induced quenching and in certain other combinations, it undergoes PSLR-induced enhancement in the ACG.

## 2.3. Optical setup for broadband FWM

To perform the broadband FWM experiment, two lasers are used to produce the three input beams. One of them is an NB source at 776 nm with a pulse duration of 2.6 ps (Toptica FemtoFiber pro NIR). The other synchronized BB laser source emits temporally ultra-compressed (with ~18 fs pulse duration) pulses with a spectral bandwidth ranging from ~930nm – 1080 nm (Toptica FemtoFiber pro UCP). All the laser beams are made circularly polarized by a Berek compensator (5540M, Newport Spectra-Physics GmbH) and a broadband quarter-wave plate (467-4215, Eksma Optics) before entering the objective (UPLFLN20X, Olympus) with a magnification of 20X and a numerical aperture (NA) of 0.5. Details of the setup can be found in Figure S1 in the Supporting Information. Figure 3a shows the spectra of the NB and BB lasers. Both operate at a 40 MHz repetition rate. The average power was approximately 30 mW and 10 mW for the NB and BB laser sources, respectively. The two laser beams were spatially and temporally combined and focused on the ACG through the objective.

The ACG for broadband FWM was raster scanned against the laser focal spot by moving it with the three-axis piezo nano-positioning stage (Nano-PDQ series, Mad City Laboratories), and the generated spectra were recorded at each excitation position by a spectrometer (Kymera 193i, Andor). This allows the construction of the hyperspectral image of the generated optical signal for the investigation of ACG's enhancement effect. Figure 3b shows the broadband FWM spectra collected from the ACG at azimuthal angles φ = 180°, 45°, and 0°. The FWM signals generated from the 2-color and 3-color schemes are on the short and long wavelength side of the spectrum, respectively. Clearly, the efficiency of FWM greatly depends on the grating periodicity. In the following, we focus on the case of 3-color FWM and use it to reveal the importance of the LSPR and PSLR in the FWM enhancement.

## 3. Results and discussion

As we can infer from Figure 3b, the broadband FWM signal within ~700 nm – 747 nm (~1500 cm$^{-1}$ – 509 cm$^{-1}$) is obtained from the 3-color scheme. Since FWM involves multiple excitation and emission wavelengths, it is not straightforward to show all the different input and output combinations involved in the process. In particular, for the 3-color scheme ($\omega_{\text{FWM}} = \omega_{\text{BB}} - \omega'_{\text{BB}} + \omega_{\text{NB}}$), where the BB laser source provides two input beams, all possible combinations of $\omega_{\text{BB}} - \omega'_{\text{BB}}$ need to be considered. For the sake of simplicity in analysis, only the experimental results of FWM at one representative output wavelength from the 3-color scheme are demonstrated here. In Figure 3b, it can be seens that the 747 nm wavelength exhibits the strongest FWM signal. Here it should be noted that, due to the dichromatic mirrors and optical filters used to isolate the generated FWM signal from the excitation wavelength, it was not possible by our setup to detect the signal at wavelengths longer than 750 nm. Even though higher efficiencies might exist in wavelength regime longer than 750 nm, it is experimentally challanging to detect the FWM signal in this range as it gets close to the excitation laser line of the NB pulse. Therefore, we investigate and analyze different contributing combinations of input wavelengths that lead to the generation of FWM output at 747 nm. For example, a combination of input wavelengths of 965 nm and 1014 nm provided by the BB source can produce the needed energy difference for the desired FWM. This, in turn, mixes with the NB laser pulse at 776 nm to produce an FWM signal at 747 nm (~509 cm$^{-1}$). The experimentally observed angular profile and the hyperspectral intensity map of the collected FWM signal at 747 nm in the ACG are shown in Figure 4a and Figure 4b, respectively. Note that in the 3-color scheme, the output at 747 nm can arise from a variety of wavelength combinations as long as the relation among the input beams is correct, i.e. $\omega_{FWM} = \omega_{BB} - \omega'_{BB} + \omega_{NB}$. A movie showing the evolution of intensity distribution with respect to the emission wavelength is provided in the Supporting Information.

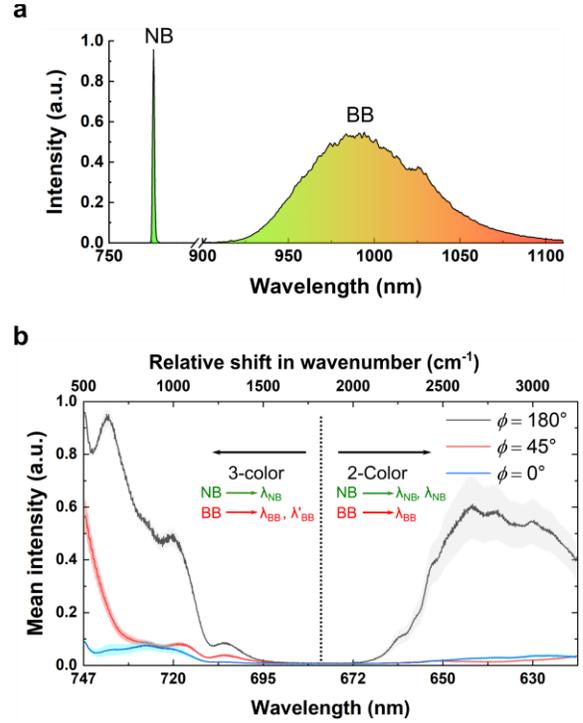

**Figure 3.** (a) The spectra of the NB and BB laser sources used in this work. (b) broadband FWM spectra from ACG at three different in-plane azimuthal angles, namely φ = 180° (black), 45° (red), and 0° (blue). The spectra to the left come from the 3-color FWM process, where the NB source provides input at the $\lambda_{\text{NB}}$, and different spectral components of the BB source offer inputs at various combinations of $\lambda_{\text{BB}}$ and $\lambda'_{\text{BB}}$. The spectra to the right are collected simultaneously from the 2-color FWM process. The shaded area indicates the standard deviation of the mean spectra obtained from different positions along the same azimuthal angle on the ACG.

The overall enhancement in a nonlinear optical process of FWM can be described by,[17]

$$\text{EF}_{\text{overall}} = \text{EF}_{\lambda_{BB}} \times \text{EF}_{\lambda'_{BB}} \times \text{EF}_{\lambda_{NB}} \times \text{EF}_{\lambda_{FWM}} \quad (4)$$

where EF denotes the enhancement factor in the intensity at the respective input and output wavelengths. Since different combinations of input wavelengths are responsible for the broadband FWM output, Equation 4 can be rewritten as follows,

$$\text{EF}_{\text{overall}} = a(\lambda_{NB}) \times \text{EF}_{\lambda_{NB}} \times \text{EF}_{\lambda_{FWM}} \times \int_{-\infty}^{\infty} b(\lambda_{BB}) \times \text{EF}_{\lambda_{BB}}(\lambda_{BB}) \times c(\lambda'_{BB}) \times \text{EF}_{\lambda'_{BB}}(\lambda'_{BB}) d\lambda_{BB} \quad (5)$$

Here, $\lambda'_{BB}$ and $\lambda_{BB}$ are paired and the requirement is that the energy difference between them should be the correct one to mix with the NB photon for the generation of the FWM signal at the desired output wavelength. To account for the intra-pulse mixing between the spectral components from the BB pulse, which has an infinite number of combinations, we select five combinations that would result in an FWM output at 747 nm within the spectral profile of the BB beam shown in Figure 3a. In Equation 5, *a* is the weight of the NB pulse used in our simulation and it has been normalized to 1, whereas, *b* and *c* are the relative weights of the different spectral components of the BB pulse in relation to *a*. These weights have been determined based on the measured power of the NB and BB pulses and the respective spectrum of the BB pulse from the experiment. The details of the relative weights of the different spectral components of the BB pulse with respect to the NB pulse are provided in the Supporting Information (Figure S2). Each combination of input consists of three wavelengths. While $\lambda_{NB}$ has been kept constant at 776 nm for every combination, the five different $\lambda_{BB}$ and $\lambda'_{BB}$ pairs that have been used for our simulation are 930 nm and 976 nm, 950 nm and 998 nm, 965 nm and 1014 nm, 980 nm and 1031 nm, and 1000 nm and 1053 nm. The pair of 965 nm and 1014 nm is particularly chosen because these two wavelengths have high and roughly equal intensity values in the BB spectrum. In essence, the range of wavelengths has been chosen in such a way that it covers the most effectively contributing span of the BB pulse. We simulate the collective enhancement in the intensity of the FWM process in ACG using FDTD (Lumerical 2020 R2.3). The simulated near-field enhancement in the intensity for the input wavelengths and the far-field enhancement in the intensity for the output wavelength have been used to calculate the overall enhancement factor for the FWM processes by employing Equations 4 and 5. The simulation considers a periodic system with periods ranging from 300 nm to 1100 nm. A p-polarized plane wave source at normal incidence has been used to simulate the excitation condition. This makes a major difference between the experimental result and the simulated one, as will be discussed later. Furthermore, the enhancement factor in gold was monitored at a point 3 nm below the bottom of the groove and at two other points 3 nm to the left and to the right of the bottom of the groove. For the emission condition, three dipole emitters were placed at the aforementioned locations and the intensity of emission at 747 nm has been integrated over a distribution of angles corresponding to NA of 0.5 (half-angle of 30°) at a height of 20 nm above the surface of the grating. The enhancement factor in the intensity for an FWM output at 747 nm from the simulation has been provided as a function of the azimuthal angle (degree) in Figure 4c. The overall enhancement in the intensity of the broadband FWM has been calculated using Equation 5 by taking into account the information gathered from the excitation and the emission simulations. The influence of the LSPR and the PSLR on the overall enhancement mechanism of broadband FWM in the non-spectral parametric/spatial domain of the ACG can be clearly seen.[35] The broader peaks at the higher azimuthal angles are attributed to the LSPR whereas the asymmetric narrower peaks in the lower azimuthal angles (around ±60º) are attributed to the PSLR. Note that the PSLR profile shown in Figure 4c has a Fano profile in the parametric domain,[35] instead of the frequency domain. The intensity enhancement factors in the parametric domain for the aforementioned wavelengths contributing to the FWM output at 747 nm have been provided along with further details pertaining to the simulation in the Figure S3 and Figure S4 of the Supporting Information.

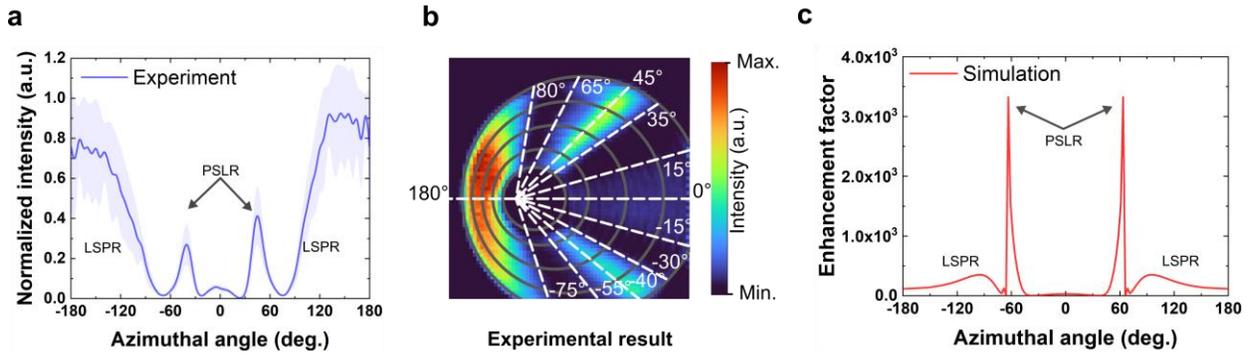

**Figure 4.** (a, b) Experimentally observed normalized broadband FWM intensity and (c) simulated broadband FWM intensity enhancement factor in the ACG at $\lambda_{FWM}$ = 747 nm (~509 cm$^{-1}$) via the 3-color scheme. The shaded area in (a) represents the standard deviation from the mean value of the experimentally recorded intensity at each azimuthal angle.

The simulation results show some deviations from the experimental results, namely, the very high quality factor of the PSLR mode around the azimuthal angle of ±60º and a comparatively lower signal at higher azimuthal angles/smaller periodicity. These deviations can be explained by considering the differences between the experimental scenario and the simulated one. For the excitation condition, it must be reiterated that in the experiment, only a limited number of grooves are illuminated at any instant at each azimuthal angle as both the ACG and the focal spot (focal diameter at an intensity of $1/e^2$ of the peak is ~1300 nm) is finite. In contrast, the simulation condition involves an infinitely extending plane wave illuminating an infinite number of grooves at each azimuthal angle. Since the number of illuminated grooves in the simulation is infinite, the simulated PSLR mode shows a peak with a much higher quality factor and amplitude compared to the experimental observation. At high azimuthal angles (lower periodicity) where the LSPR dominates, the efficiency of hot spot generation and consequently its strength is higher in comparison to that in the PSLR region of the ACG. This is because the number of grooves illuminated by the laser beam is higher in the region dominated by the LSPR compared to the region where the PSLR dominates. A similar effect also originates in emission detection with a finite size of spectrometer slit and a given magnification of the optical setup. These effects might scale the resultant broadband FWM signal in the experiment with the factor that is a function of the grating period with more signal coming from the higher azimuthal angles. Other contributions may come from the difference between the simulated and the experimental groove geometries, as well as the uncertainty in the relative weights of the NB pulse and the different spectral components of the BB pulse. Furthermore, the movement of the detection point in rectangular coordinates to measure the signal from the ACG also gives rise to the Moiré effect in the detected broadband FWM signal. Nevertheless, the simulated results in Figure 4c could well reproduce most of the spectral features observed in a very complex nonlinear experimental phenomenon of FWM in a reasonably accurate manner, and describe all the modes of the grating involved in generating broadband FWM from an ACG.

## 4. Conclusions

In this work, we have demonstrated broadband FWM in a gold grating and shown that the LSPR and PSLR play a major role in the enhancement of the FWM signal. Considering the capability for broadband enhancement of the ACG, this work also demonstrates the ACG's applicability in broadband nonlinear optical processes. The broadband FWM signal obtained from the 3-color scheme is in the molecular fingerprint region (500 cm$^{-1}$ – 1500 cm$^{-1}$). Therefore, it is a major noise source for plasmonic SECARS in this spectral regime. Our study shows that the non-resonant noise of SECARS is intrinsic and can be enhanced by the structures. Further strategies should be developed to separate such spectrally overlapping noise from the FWM of the plasmonic nanostructures. The developed optical setup with the capability of performing broadband FWM experiments in conjunction with the ACG's capability for broadband surface enhancement paves the way for investigations like broadband SECARS into molecular systems ranging from monolayers to the single-molecular limit.

## ASSOCIATED CONTENT

### SUPPORTING INFORMATION

Experimental setup, NB and BB pulse spectra, FDTD simulation: FWM, Evolution of intensity distribution with the variation in wavelength (video).

## AUTHOR INFORMATION

### Author Contribution note

The authors Abhik Chakraborty and Parijat Barman have contributed equally to this work.

### Corresponding Authors

**Jer-Shing Huang** – Leibniz Institute of Photonic Technology, 07745 Jena, Germany; Institute of Physical Chemistry and Abbe Center of Photonics, Friedrich Schiller University Jena, D-07743 Jena, Germany; Research Center for Applied Sciences, Academia Sinica, Taipei 11529, Taiwan; Department of Electrophysics, National Yang Ming Chiao Tung University, Hsinchu 30010, Taiwan;
Email: jer-shing.huang@leibniz-ipht.de

**Jürgen Popp** – Leibniz Institute of Photonic Technology, 07745 Jena, Germany; Institute of Physical Chemistry and Abbe Center of Photonics, Friedrich Schiller University Jena, D-07743 Jena, Germany;
Email: juergen.popp@leibniz-ipht.de

## ACKNOWLEDGMENT

We would like to acknowledge the financial support from DFG via CRC 1375 NOA (398816777). Furthermore, the funding from the European Union's Horizon 2020 research and innovation programme under grant agreement No 101016923 is gratefully acknowledged. The authors are grateful to Oliver Rüger and Yi-Ju Chen for their support during the fabrication of samples and the analysis of data, respectively.

## ABBREVIATIONS

ACG, azimuthally chirped grating; FWM, four-wave mixing; SECARS, surface-enhanced coherent anti-Stokes Raman scattering; SPP, surface plasmon polariton; LSPR, localized surface plasmon resonance; PSLR, plasmonic surface lattice resonance; BB, broadband; NB, narrowband; FDTD, finite-difference time-domain.

## *Supporting information*

# Broadband Four-Wave Mixing Enhanced by Plasmonic Surface Lattice Resonance and Localized Surface Plasmon Resonance in an Azimuthally Chirped Grating

Abhik Chakraborty,[†,‡] Parijat Barman,[†,‡] Ankit Kumar Singh,[‡] Xiaofei Wu,[‡] Denis A. Akimov,[‡] Tobias Meyer-Zedler,[†,‡] Stefan Nolte,[||, §] Carsten Ronning,[⊥] Michael Schmitt,[†] Jürgen Popp,[†,‡] Jer-Shing Huang[†,‡,#,††]

†Institute of Physical Chemistry and Abbe Center of Photonics, Friedrich-Schiller-Universität Jena, Helmholtzweg 4, D-07743 Jena, Germany

‡Leibniz Institute of Photonic Technology, Albert-Einstein Str. 9, 07745 Jena, Germany

||Institute of Applied Physics, Abbe Center of Photonics, Friedrich-Schiller-Universität Jena, Albert-Einstein-Straße 15, Jena 07745, Germany

§Fraunhofer Institute for Applied Optics and Precision Engineering IOF, Center of Excellence in Photonics, Albert-Einstein-Str. 7, Jena 07745, Germany

⊥Institut für Festkörperphysik, Friedrich-Schiller-Universität Jena, Max-Wien-Platz 1, 07743 Jena, Germany

#Research Center for Applied Sciences, Academia Sinica, 128 Sec. 2, Academia Road, Nankang District, Taipei 11529, Taiwan

††Department of Electrophysics, National Yang Ming Chiao Tung University, Hsinchu 30010, Taiwan

**Content:**

I. Experimental setup (Figure S1)

II. NB and BB pulse spectra (Figure S2)

III. FDTD simulation

IV. Evolution of intensity distribution with the variation in wavelength (video)

## I. Experimental setup

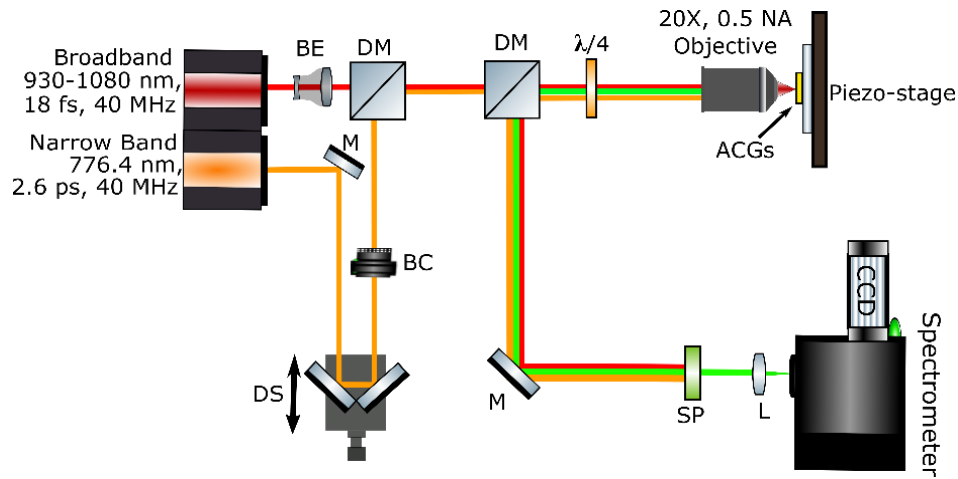

**Figure S1.** (a) Setup for BFWM experiment. BE: beam expander, DM: dichroic mirror, M: mirror, λ/4: quarter wave-plate, SP: short-pass filter, L: lens, DS: delay stage, BC: Berek compensator.

## II. NB and BB pulse spectra

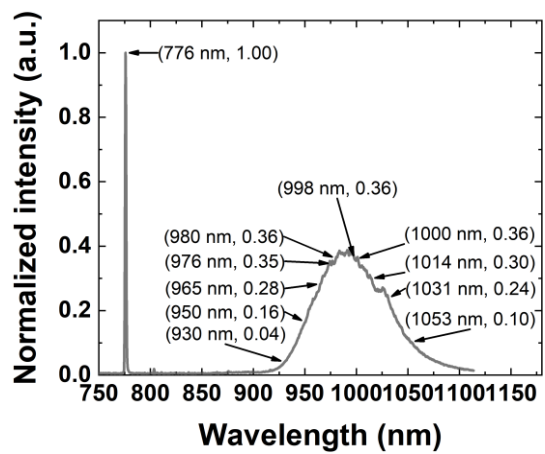

**Figure S2:** Spectra of the NB pulse and the BB pulse. The weights of the representative spectral components of the BB pulse have been calculated in relation to the normalized weight (1) of the NB pulse. The respective spectral components and their corresponding weights have been annotated within the plot. These weights have been used in conjunction with the simulated data to calculate the overall intensity enhancement factor of the FWM signal.

## III. FDTD simulation: FWM

Incidence Condition

A plane wave at normal incidence has been used to excite an infinitely extending gold grating. A 2D simulation region with periodic boundary condition and a mesh resolution of 1 nm × 1 nm has been employed in the simulation. The polarization of the incidence is oriented along the width of the grooves of the grating. Three point monitors have been positioned strategically near the bottom of the groove to record the intensity enhancement in the region that contributes the most to the enhancement of the nonlinear optical process. One point monitor has been positioned 3 nm below the groove and the other two have been positioned 3 nm to the left and 3 nm to the right of the bottom of the groove. Figure S3a demonstrates the simulation setup for the incidence condition. To account for the varying periodicity of the ACG, the periodicity in the simulation has been varied from 300 nm (azimuthal angle of 180°) to 1100 nm (azimuthal angle of 0°) at a constant increment of 20 nm. Considering the structural symmetry of the ACG on either side of the horizontal axis, the simulated data between 0° and 180° is essentially a mirror image of the one between 0° and -180°. Since 3-color FWM with an output at 747 nm can be an outcome of different combinations of wavelengths in our broadband FWM setup, the intensity enhancement has been recorded for the NB pulse (776 nm) and at the different representative pairs of spectral components from the BB pulse (930 nm and 976nm, 950 nm and 998 nm, 965 nm and 1014 nm, 980 nm and 1031 nm, 1000 nm and 1053 nm). Figure S3b to Figure S3g show the intensity enhancement over the whole range of azimuthal angles of the ACG for these different spectral components of the incidence condition.

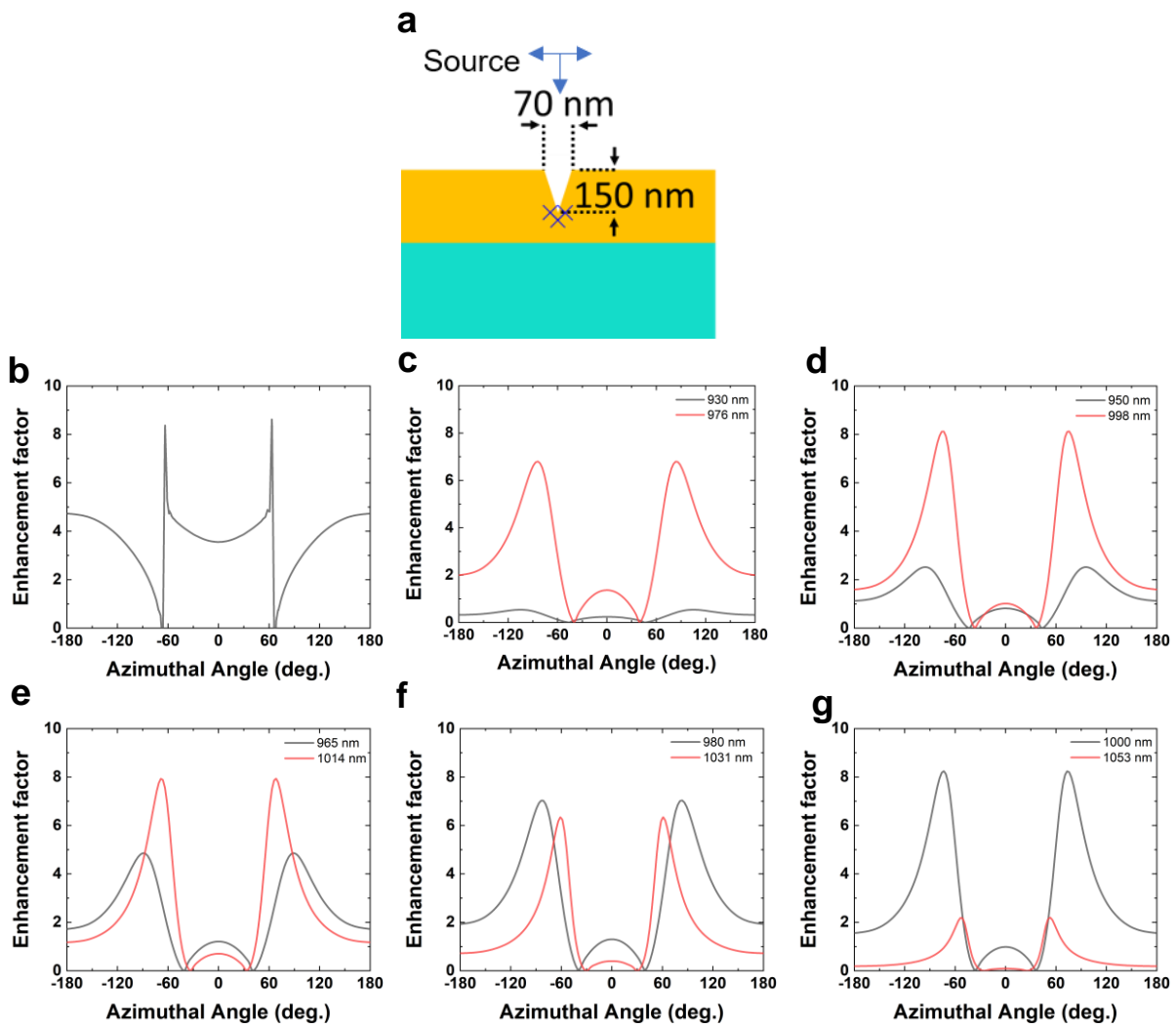

**Figure S3:** a) The setup to simulate the incidence condition has been shown. Each period of the gold ACG has a V-shaped groove of 150 nm depth and 70 nm opening width. The three point monitors are denoted by the symbol × and they have been kept 3 nm below, 3 nm to the left and 3 nm to the right of the bottom of the groove. A plane wave incidence with p-polarization has been used at normal incidence to excite the system. The weighted intensity enhancement factors simulated at the wavelengths b) 776 nm (NB pulse), c) 930 nm and 976 nm, d) 950 nm and 998 nm, e) 965 nm and 1014 nm, f) 980 nm and 1031 nm, g) 1000 nm and 1053 nm have been shown.

Emission Condition

To simulate the emission condition, the structure as well as other parameters like the boundary condition and the mesh resolution have been kept the same as they were in the incidence condition. However, the 3 point monitors from the incidence condition have been replaced by 3 dipole emitters in this case. A line monitor positioned 20 nm above the grating surface has been used to integrate the emission intensity at 747 nm over a distribution of angles corresponding to an NA of 0.5 (half-angle of 30°). The simulation setup has been shown in Figure S4a. In order to calculate the intensity enhancement factor in the emission condition, the integrated emission intensity for each azimuthal angle of the ACG has been normalized by the integrated emission intensity collected from three dipole emitters in vacuum which are kept at exactly the same position as in the ACG. The calculated enhancement factor over the whole range of azimuthal angles of the ACG for the emission condition has been demonstrated in Figure S4b.

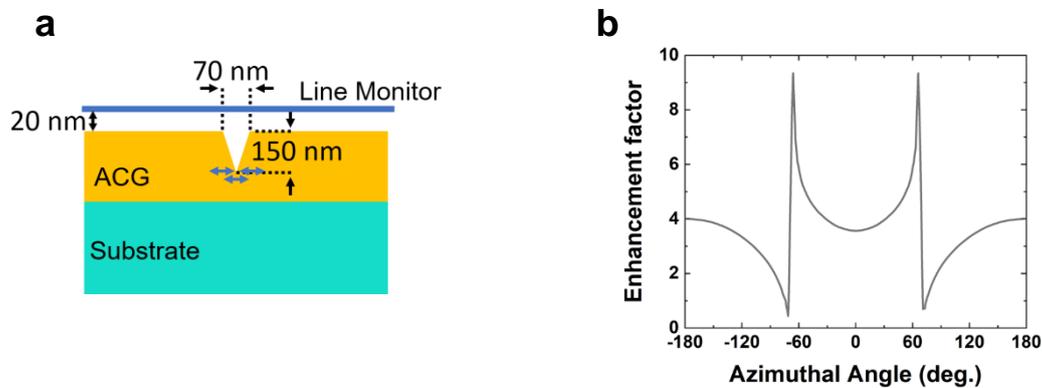

**Figure S4:** a) The setup to simulate the emission condition at a wavelength of 747 nm for each period has been demonstrated. The three dipole monitors are denoted by the symbol ↔ and they have been kept 3 nm below, 3 nm to the left and 3 nm to the right of the bottom of the groove. The intensity of emission has been integrated over a distribution of angles corresponding to NA of 0.5 (half-angle of 30°) at the line monitor situated 20 nm above the grating. b) The intensity enhancement factor for the far-field emission at 747 nm.

### IV. Video of BFWM signal of ACG

''ACG_BFWM.avi''

The movie shows the BFWM signal distribution on ACG for relative anti-Stokes shift from~509 cm$^{-1}$ to 3000 cm$^{-1}$.